\documentclass[fleqn,twoside,onecolumn,nofootinbib]{revtex4}

\usepackage{amsmath,amssymb,graphicx}

\begin{document}

\title{Creation of bielectron of Dirac cone: the tachyon solution in magnetic field.}

\author{Lyubov E. Lokot}\email{Corresponding_author: llokot@gmail.com, tel: +380509030899, fax: +380445256033}

\affiliation{Institute of Semiconductor Physics, NAS of Ukraine, 41, Nauky Ave., Kyiv 03028, Ukraine}

\begin{abstract}
Schr\"{o}dinger equation for pair of two massless Dirac particles when magnetic field is applied in Landau gauge is solved exactly. In this case the separation of center of mass and relative motion is obtained. Landau quantization $\epsilon=\pm\,B\sqrt{l}$ for pair of two Majorana fermions coupled via a Coulomb potential from massless chiral Dirac equation in cylindric coordinate is found. The root ambiguity in energy spectrum leads into Landau quantization for bielectron, when the states in which the one simultaneously exists are allowed. The tachyon solution with imaginary energy in Cooper problem ($\epsilon^{2}<0$) is found. The continuum symmetry of Dirac equation allows perfect pairing between electron Fermi spheres when magnetic field is applied in Landau gauge creating a Cooper pair.
\end{abstract}

\maketitle

\section{Introduction}

The graphene ~\cite{{Novoselov},{Vasko},{Zhang}} presents a new state of matter of layered materials. The energy bands for graphite was found using "tight-binding" approximation by P.R. Wallace ~\cite{{Wallace}}. In the low-energy limit the single-particle spectrum is Dirac cone similarly to the light cone in relativistic physics, where the light velocity is substituted by the Fermi velocity $v_{F}$ and describes by the massless Dirac equation.

The graphene is the single graphite layer, i. e. two-dimensional graphite plane of thickness of single atom. The graphene lattice resembles a honeycomb lattice. The graphene lattice one can consider like into the composite of two triangular sublattices. In 1947 Wallace in "tight-binding" approximation consider a graphite which consist off the graphene blocks with taken into account the overlap only the nearest $\pi$-electrons.

The two-dimensional nature of graphene and the space and point symmetries of graphene acquire of the reason for the massless electron motion since lead into massless Dirac equation (Majorana fermions) ~\cite{{Wallace},{Semenoff}}. At low-energy limit the single particle spectrum forms with $\pi$-electron carbon orbital and consist off completely occupation valence cone and completely empty conduction cone, which have cone like shape with single Dirac point. In Dirac point the existing an electron as well as a hole is proved. The state in Dirac cone is double degenerate with taken into account a spin.

The existing of the massless Dirac fermions in graphene was proved based on the unconventional quantum Hall effect. The reason of creation the integer Hall conductivity ~\cite{{Zheng},{Gusynin1},{Gusynin2},{Peres}} is derived from Berry phase ~\cite{{Berry},{Carmier}}.

When the magnetic field is applied perpendicularly into graphene plane the lowest (n=0) Landau level has the energy $\pm\Delta$ in two nonequivalent cones $K_{\mp}$, correspondingly ~\cite{Gusynin3}. In the paper ~\cite{Gusynin3} the Dirac mass via a splitting value is found when Zeeman coupling is absence. These properties of the lowest Landau level which distribute between particles and antiparticles in equal parts are base of the integer quantum Hall effect in graphene ~\cite{Gusynin3}. For $n\geq1$ an all Landau levels are fourfold degenerate. For $n=0$ a states in both cones are twofold degenerate with energies $\pm\Delta$ with taken into account a spin ~\cite{Gusynin3}.

\section{Solution of massless chiral Dirac equation for pair of two Majorana fermions coupled via a Coulomb potential in magnetic field in Landau gauge.}

Calculation model of the graphene reflects continuum symmetry of QED$_{2+1}$ including Lorentz group. SU(2) symmetry are shown to be found similar to chiral in the paper ~\cite{Gusynin3}. Hence is conserving quantum number of chirality.

The energy bands for graphene was found using "tight-binding" approximation in the papers ~\cite{{Wallace},{Gusynin3}}.

Calculate the quantized Landau energy as well as the wave function of the Majorana particles in cylindrical coordinate in magnetic field in Landau gauge. Enter the production and annihilation operators as following:
\begin{equation}\label{deq0}
\begin{array}{cccc}
\hat{c}^{\dag}=\sqrt{B}(-\frac{\partial}{\partial\,\xi_{2}}+\xi_{1}),\\
\hat{c}=\sqrt{B}(\frac{\partial}{\partial\,\xi_{1}}+\xi_{2}),\\
\end{array}
\end{equation}
where $\xi_{1}=\frac{\sqrt{B}}{2}\zeta$, $\xi_{2}=\frac{\sqrt{B}}{2}\eta$, $\zeta=x+iy$, $\eta=x-iy$, which satisfies the commutator relation:
\begin{equation}
[\hat{c}^{\dag},\hat{c}]=2B.
\end{equation}

Hence for noninteracting Dirac particles we write the massless Dirac equation in the form:

\begin{equation}
\frac{i}{\sqrt{2}}\hbar\,v_{F}\left\|
\begin{array}{cc}
0 & \hat{c}^{\dag}\\
\hat{c} & 0 \\
\end{array}
\right\|\left\|
\begin{array}{cc}
\Psi_{1}\\
\Psi_{2}\\
\end{array}
\right\|=\varepsilon\left\|
\begin{array}{cc}
\Psi_{1}\\
\Psi_{2}\\
\end{array}
\right\|.\end{equation}

The Schr\"{o}dinger equation for the reduced energy can be rewritten in the form:

\begin{equation}
\frac{1}{2}\hat{c}^\dag\,\hat{c}\,\Psi=\epsilon^{2}\Psi.
\end{equation}

For graphene in vacuum the effective fine structure parameter $\alpha_{G}=\frac{e^{2}}{v_{F}\hbar\kappa\sqrt{\pi}}=1.23$. For graphene in substrate $\alpha_{G}=0.77$, when the permittivity of graphene in substrate is estimated to be $\kappa=1.6$ ~\cite{{Alicea}}. It means the prominent Coulomb effects. Hence the Coulomb potential may be found in the form ~\cite{Novikov}:
\begin{equation}\label{deq01}
V(\rho)=\hbar\,v_{F}\frac{\alpha}{\rho},
\end{equation}

where $v_{F}=10^{6}$ m/s is the graphene Fermi velocity (here we assume that $\hbar=1$).

\begin{figure}
\includegraphics*[bb=5 10 1000 600,width=5in]{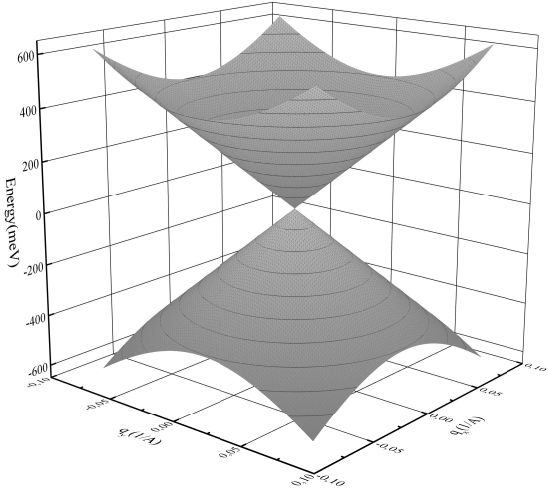}
\caption{(Color online) Single-particle spectrum of graphene for massless Dirac fermions (Majorana fermions).}
\end{figure}

When magnetic field is applied perpendicularly into graphene plane in z axis along field distribution. The vector potential in the gauge ~\cite{Landau} $\textbf{A}=\frac{1}{2}[\textbf{Hr}]$ has a components $A_{\varphi}=H\rho/2$, $A_{\rho}=A_{z}=0$ and Schr\"{o}dinger equation:
\begin{equation}\label{deq1}
\begin{array}{cccc}
-4\frac{\partial^{2}\Psi}{\partial\,\rho^{2}}-\frac{4}{\rho}\frac{\partial\Psi}{\partial\rho}-\frac{4}{\rho^{2}}\frac{\partial^{2}\Psi}{\partial\varphi^{2}}-\frac{4i}{\rho^{2}}\frac{\partial\Psi}{\partial\varphi}
-2iB\frac{\partial\Psi}{\partial\varphi}+\frac{B^{2}}{4}\rho^{2}\Psi-2B\Psi=2\epsilon^{2}\Psi.
\end{array}
\end{equation}

The Schr\"{o}dinger equation \eqref{deq1} with including the Coulomb potential Eq. \eqref{deq01} one can rewritten in the form:

\begin{equation}\label{deq*}
\begin{array}{cccc}
-4\frac{\partial^{2}\Psi}{\partial\,\rho^{2}}-\frac{4}{\rho}\frac{\partial\Psi}{\partial\rho}-\frac{4}{\rho^{2}}\frac{\partial^{2}\Psi}{\partial\varphi^{2}}-\frac{4i}{\rho^{2}}\frac{\partial\Psi}{\partial\varphi}
-2iB\frac{\partial\Psi}{\partial\varphi}+\frac{B^{2}}{4}\rho^{2}\Psi-2B\Psi+\frac{\alpha^{2}}{\rho^{2}}\Psi=2\epsilon^{2}\Psi.
\end{array}
\end{equation}

The solution Eq. \eqref{deq*} with including the Coulomb potential Eq. \eqref{deq01} can look for in the form:
\begin{equation}
\Psi_{n,k}=\frac{1}{\sqrt{2\pi}}R(\rho)e^{im\varphi}.
\end{equation}

Substituting the solution in Eq. \eqref{deq*}, one can find for the radial function the following equation
\begin{equation}
R''+\frac{1}{\rho}R'+(\beta-\frac{k^{2}}{\rho^{2}}-2\gamma\,m-\gamma^{2}\rho^{2})R=0,
\end{equation}
where $\beta=(\epsilon^{2}+B)/2$, $\gamma=B/4$, $k^2=m^2+m-\alpha^2$.
Entering the new independent variable $\xi=\gamma\rho^2$, the equation \eqref{deq*} can be rewritten in the form:
\begin{equation}\label{deq2}
\xi\,R''+R'+(\frac{\xi}{4}+\lambda-\frac{k^{2}}{4\xi})R=0,
\end{equation}
where $\lambda=\frac{\beta}{4\gamma}-\frac{m}{2}$.
At $\xi\rightarrow\infty$ conduct of sought for function are shown to be found as following $e^{-\xi/2}$, and at $\xi\rightarrow0$ like $\xi^{k/2}$.

The solution of the Eq. \eqref{deq2} can look for in the form:
\begin{equation}\label{deq3}
R=e^{-\xi/2}\xi^{k/2}\varpi(\xi).
\end{equation}

To substitute the solution \eqref{deq3} in the Eq. \eqref{deq2} it is necessarily to find as follows
\begin{equation}\label{deq4}
R''=\varpi''e^{-\xi/2}\xi^{k/2}+2\varpi'\frac{\partial}{\partial\,\xi}e^{-\xi/2}\xi^{k/2}+\varpi\frac{\partial^{2}}{\partial\,\xi^{2}}e^{-\xi/2}\xi^{k/2},
\end{equation}
\begin{equation}\label{deq5}
R'=\varpi'e^{-\xi/2}\xi^{k/2}+\varpi\frac{\partial}{\partial\,\xi}e^{-\xi/2}\xi^{k/2}.
\end{equation}
Since
\begin{equation}\label{deq6}
\frac{\partial}{\partial\,\xi}e^{-\xi/2}\xi^{k/2}=-\frac{1}{2}e^{-\xi/2}\xi^{k/2}+\frac{k}{2}e^{-\xi/2}\xi^{k/2-1},
\end{equation}
\begin{equation}\label{deq7}
\begin{array}{cccc}
\frac{\partial^{2}}{\partial\,\xi^{2}}e^{-\xi/2}\xi^{k/2}=\frac{1}{4}e^{-\xi/2}\xi^{k/2}-\frac{k}{2}e^{-\xi/2}\xi^{k/2-1}+\frac{k}{2}(\frac{k}{2}-1)e^{-\xi/2}\xi^{k/2-2}.\\
\end{array}
\end{equation}
Substituting \eqref{deq4}, \eqref{deq5}, \eqref{deq6}, \eqref{deq7} into the Eq. \eqref{deq2} we find the equation for $\varpi(\xi)$ as follows
\begin{equation}\label{deq8}
\xi\varpi''+\varpi'(1+k-\xi)+\varpi(\lambda-\frac{k+1}{2})=0.
\end{equation}

Hence for $\varpi(\xi)$ we derive the equation for confluent hypergeometric function:
\begin{equation}
\varpi=F(-(\lambda-\frac{k+1}{2}),k+1,\xi).
\end{equation}
From the condition of finite of the wave function one can find the energy spectrum in the form:
\begin{equation}
\epsilon=\pm\,B\sqrt{l},
\end{equation}
where $l=2n+m+k$, $\lambda-\frac{k+1}{2}=n$,  $m=0,1,2,3,...$. The wave function expressed via the associated Laguerre polynomial:
\begin{equation}
R_{n,k}=[\frac{B(n-k)!}{2n!^{3}(n-k+1)}]^{1/2}\psi_{n,k},
\end{equation}
where
\begin{equation}
\psi_{n,k}=e^{-\frac{\xi}{2}}\xi^{\frac{k}{2}}L_{n}^{k}(\xi).
\end{equation}

Because the solution for the wave functions for the pair of two massless Dirac particles when magnetic field is applied in Landau gauge one can express via the product of the two identical wave functions one can conclude that in this case the separation of center of mass and relative motion is shown ~\cite{lol}.

Entering the production and annihilation operators as following \eqref{deq0} and solving Schr\"{o}dinger equation one can derive the known for quantum electrodynamics (QED) solution - the root ambiguity in energy spectrum $\epsilon=\pm\,B\sqrt{l}$, where $l$ is a number of natural numbers set ~\cite{McClure}. The root ambiguity in energy spectrum at the solution of the problem about quantization with relativistic invariance lead in quantum field theory into the creation of a pairs of particles (particles+antiparticles) ~\cite{Berestetskii}. When $l$ is a number of complex numbers set the tachyon solutions are provided by arising the complex energy in spectrum of quantization of Landau for pair of two Majorana fermions coupled via a Coulomb potential.

For graphene with strong Coulomb interaction the Bethe-Salpeter equation for the electron-hole bound state was solved and a tachyonic solution was found ~\cite{Gamayun}.

Calculation model of the graphene reflects continuum symmetry of QED$_{2+1}$ including Lorentz group. SU(2) symmetry are shown to be found similar chiral in the paper ~\cite{Gusynin3}. Hence is conserving quantum number of chirality. In the paper ~\cite{Malard} the selection rules for the electron-radiation and for the electron-phonon interactions at all points in the Brillouin zone are derived based on irreducible representation of the crystallographic space groups as well as the symmetry properties of electrons and phonons. The each of these models are qualitatively different.

In the paper ~\cite{Nandkishore} a chiral superconductivity from electron-electron repulsive in doped graphene in the $M$ point is predicted.

In the paper ~\cite{Rashba} a Magneto-Coulomb levels at a three-dimensional saddle point were found. The Schr\"{o}dinger equation for the three-dimensional saddle surface  geometry at the magnetic field is applied unconventionally was solved exactly in the paper ~\cite{Rashba} by reducing into one-dimensional Schr\"{o}dinger equation.

In the paper ~\cite{Hartmann} the exciton binding energy is scaled with the formed band gap when the magnetic field is applied and an exciton insulator transition in carbon nanotubes was not found and their THz application was predicted.

In the paper ~\cite{Julian} in the UCoGe material the high-temperature superconductivity is connected with spin fluctuations and hence may be reduced by magnetic field is applied.

The exciton Wannier equation for graphene was solved in the papers ~\cite{{lokot},{Stroucken},{Stroucken1},{Malic}}. A very large exciton binding energies were found. In the paper ~\cite{lokot} a theoretical study the both the quantized energies of excitonic states and their wave functions in graphene is presented. An integral two-dimensional Schr\"{o}dinger equation of the electron-hole pairing for a particles with electron-hole symmetry of reflection is exactly solved. The solutions of Schr\"{o}dinger equation in momentum space in graphene by projection the two-dimensional space of momentum on the three-dimensional sphere are found exactly. We analytically solve an integral two-dimensional Schr\"{o}dinger equation of the electron-hole pairing for particles with electron-hole symmetry of reflection. In single-layer graphene (SLG) the electron-hole pairing leads to the exciton insulator states. Quantized spectral series and light absorption rates of the excitonic states which distribute in valence cone are found exactly. If the electron and hole are separated, their energy is higher than if they are paired. The particle-hole symmetry of Dirac equation of layered materials allows perfect pairing between electron Fermi sphere and hole Fermi sphere in the valence cone and conduction cone and hence driving the Cooper instability.

\section{Conclusions}

Schr\"{o}dinger equation for pair of two massless Dirac particles when magnetic field is applied in Landau gauge is solved exactly. Landau quantization $\epsilon=\pm\,B\sqrt{l}$ for pair of two Majorana fermions coupled via a Coulomb potential from massless chiral Dirac equation in cylindric coordinate is found. In this case the separation of center of mass and relative motion is derived. The root ambiguity in energy spectrum leads into Landau quantization for bielectron, when the states in which the one simultaneously exists are allowed. The tachyon solution with imaginary energy in Cooper problem ($\epsilon^{2}<0$) is found. The wave function are shown to be expressed via the associated Laguerre polynomial. In the paper the Cooper problem in superconductor theory is solved as quantum-mechanical problem for two electrons unlike from the paper ~\cite{Gamayun} where the Bethe-Salpeter equation was solved for electron-hole pair. The continuum symmetry of Dirac equation allows perfect pairing between electron Fermi spheres and hence creating a Cooper pair.

\section{Mathematical Appendix}

From a algebraic manipulation one can find a following recurrence relations:

\begin{equation}
\begin{array}{cccc}
G(\alpha,\alpha-\gamma+1,-z)=(F(\alpha,\gamma,z)\frac{\Gamma(\alpha)}{\Gamma(\gamma)}z^{\gamma}-F(\alpha-\gamma+1,2-\gamma,z)\frac{\Gamma(\alpha-\gamma+1)}{\Gamma(2-\gamma)}z)\times\\
\times\frac{\Gamma(1-\alpha)\Gamma(\gamma-\alpha)}{\Gamma(\alpha)\Gamma(1-\alpha)-\Gamma(\gamma-\alpha)\Gamma(\alpha-\gamma+1)(-1)^{\gamma-1}}(-1)^{\alpha}z^{\alpha-\gamma},
\end{array}
\end{equation}
where
\begin{equation}
G(\alpha,\beta,z)=\frac{\Gamma(1-\beta)}{2\pi\,i}\int_{C_{1}}(1+\frac{t}{z})^{-\alpha}t^{\beta-1}e^{t}dt,
\end{equation}

\begin{equation}
F(\alpha,\beta,\gamma,z)=-z^{-\beta}\frac{\Gamma(\gamma)\Gamma(1-\alpha)}{\Gamma(\gamma-\beta)\Gamma(\beta+1-\alpha)}F(\beta,\beta+1-\gamma,\beta+1-\alpha,\frac{1}{z}),
\end{equation}

\begin{equation}
F(\alpha,\beta,\gamma,z)=-z^{-\alpha}\frac{\Gamma(\gamma)\Gamma(1-\beta)}{\Gamma(\gamma-\alpha)\Gamma(\alpha+1-\beta)}F(\alpha,\alpha+1-\gamma,\alpha+1-\beta,\frac{1}{z}),
\end{equation}

\begin{equation}
\begin{array}{cccc}
G(\alpha,\alpha-\gamma+1,-z)=\frac{\Gamma(1-\alpha)\Gamma(\gamma-\alpha)}{\Gamma(\alpha)\Gamma(1-\alpha)-\Gamma(\gamma-\alpha)\Gamma(\alpha-\gamma+1)z^{-2\alpha}}\times\\
\times(\frac{\Gamma(\alpha)}{\Gamma(\gamma)}z^{\gamma-\alpha}F(\alpha,\gamma,z)-\frac{\Gamma(\alpha-\gamma+1)}{\Gamma(2-\gamma)}z^{1-2\alpha}F(\alpha-\gamma+1,2-\gamma,z))e^{-z}z^{2\alpha+\gamma}(-1)^{\alpha},
\end{array}
\end{equation}

\begin{equation}
F(\alpha-\gamma+1,2-\gamma,z)=(-1)^{\alpha-\gamma+1}e^{z}\frac{\Gamma^{2}(\gamma-\alpha)}{\Gamma^{2}(1-\alpha)}F(\alpha,\gamma,-z).
\end{equation}

From a algebraic manipulation one can find a following integrals and recurrence relations which connect theirs:

\begin{equation}
\begin{array}{cccc}
J=\int_{0}^{\infty}e^{-\lambda\,z}z^{\gamma-1}F(-n,\gamma,kz)F(\alpha',\gamma,k'z)dz=\\
\Gamma^{2}(\gamma)\lambda^{\alpha'-\gamma-n}(\lambda-k)^{n}(\lambda-k')^{-\alpha'}\times\\
\times\,(F(\alpha',-n,\gamma,\frac{kk'}{(\lambda-k)(\lambda-k')})-\frac{2k'}{\lambda-k'}F(\alpha'+1,-n,\gamma,\frac{kk'}{(\lambda-k)(\lambda-k')})),\\
\end{array}
\end{equation}

\begin{equation}
\begin{array}{cccc}
J_{\nu}^{s,p}(\alpha,\alpha')=\int_{0}^{\infty}e^{-\frac{k+k'}{2}z}z^{\nu-1+s}F(\alpha,\gamma,kz)F(\alpha',\gamma-p,k'z)dz,\\
\end{array}
\end{equation}

\begin{equation}
\begin{array}{cccc}
J_{\nu}^{s,p}(\alpha,-n)=(-\frac{1}{2\pi\,i})\frac{\Gamma(1-\alpha)\Gamma(\gamma)}{\Gamma(\gamma-\alpha)}\Gamma(\nu+s)\frac{1}{\gamma(\gamma+1)...(\gamma+n-1)}(-1)^{\nu+s-\gamma+p}\frac{n!}{l!(n-l)!}\times\\
\times\,(\gamma-p+n-1)!(\nu+s-\gamma+p)!k'^{l}\lambda^{\gamma}(\lambda-k)^{-\alpha}(\lambda-k')^{-\alpha}\times\\
\times\,(F(\alpha,\alpha,\gamma,\frac{kk'}{(\lambda-k)(\lambda-k')})-\frac{2k}{\lambda-k}F(\alpha+1,\alpha,\gamma,\frac{kk'}{(\lambda-k)(\lambda-k')})),\\
\end{array}
\end{equation}

\begin{equation}
\begin{array}{cccc}
J_{\gamma}^{s,p}(\alpha,\alpha')=\int_{0}^{\infty}e^{-\frac{k+k'}{2}z}z^{\gamma-1+s}F(\alpha,\gamma,kz)F(\alpha',\gamma-p,k'z)dz,\\
\end{array}
\end{equation}

\begin{equation}
\begin{array}{cccc}
J_{\gamma}^{s,p}(\alpha,\alpha')=\frac{\gamma-1}{\gamma-\alpha-1}J_{\gamma-1}^{s+1,p-1}(\alpha,\alpha')-\frac{\alpha}{\gamma-\alpha-1}J_{\gamma}^{s,p}(\alpha+1,\alpha'),\\
\end{array}
\end{equation}

\begin{equation}
\begin{array}{cccc}
(\gamma-\alpha-1)J_{\gamma}^{s,p}(\alpha,\alpha')=(\gamma-1)J_{\gamma-1}^{s+1,p-1}(\alpha,\alpha')-\alpha\,J_{\gamma}^{s,p}(\alpha+1,\alpha'),\\
\end{array}
\end{equation}

\begin{equation}
\begin{array}{cccc}
J_{\gamma}^{s,0}(\alpha,\alpha')=\Gamma(\gamma+s)k'^{-\alpha'}(\frac{k+k'}{2})^{-\gamma-s+\alpha'}\frac{(-1)^{\alpha'-\gamma}(\alpha'-\gamma)!(\gamma+s-1)!}{\gamma(\gamma+1)...(\gamma+s-1)}F(\alpha,\gamma+s-\alpha',\gamma,\frac{2k}{k+k'}).
\end{array}
\end{equation}

\end{document}